\documentstyle[11pt]{article}
\newcommand{\bigzerou}{%
\smash{\lower1.7ex\hbox{\bg 0}}}
\newcommand{\half}{\frac{1}{2}}

\newcommand{\nonum}{\nonumber}
\newcommand{\beq}{\begin{equation}}
\newcommand{\enq}{\end{equation}}

\newcommand{\mapright}[1]{%
\smash{\mathop{%
\hbox to 1.0cm{\rightarrowfill}}\limits^{#1}}}
\newcommand{\mapleft}[1]{%
\smash{\mathop{%
\hbox to 1.3cm{\leftarrowfill}}\limits^{#1}}}

\newcommand{\no}{\nonumber}

\newtheorem{thm}{Theorem}
\newtheorem{conj}{Conjecture}
\newtheorem{notn}{Notation}
\newtheorem{cor}{Corollary}
\newtheorem{rem}{Remark}

\begin{document}

%%%%%%%%%%%%%%%%%%%title%%%%%%%%%%%
%%%%%%%%%%%%%%%%%%%%%%%%%%%%%%%%%%%

\begin{titlepage}
\vglue 3cm

\begin{center}
\vglue 0.5cm
{\Large\bf ${\cal N}=4$ Supersymmetric Yang-Mills Theory
\\ on Orbifold-$T^4/{\bf Z}_2$: Higher Rank Case}
\vglue 1cm
{\large Masao Jinzenji${}^\dagger $,  Toru Sasaki${}^* $} 
\vglue 0.5cm
{\it ${}^\dagger $
Division of Mathematics, Graduate School of Science,
        Hokkaido University, Sapporo 060-0810, Japan 
}\\
{\it ${}^*$Department of Physics,
Hokkaido University, Sapporo 060-0810, Japan}
{\it ${}^\dagger $jin@math.sci.hokudai.ac.jp \\
${}^*$ sasaki@particle.sci.hokudai.ac.jp}

\baselineskip=12pt

\vglue 1cm
\begin{abstract}
  We derive the partition function of ${\cal N}=4$ 
supersymmetric Yang-Mills theory
on orbifold-$T^4/{\bf Z}_2 $ for gauge group $SU(N)$.
We generalize the method of our previous work for the $SU(2)$ case
to the $SU(N) $ case.
The resulting partition function is  represented as the sum of 
the  product of G\"ottche formula of 
singular quotient space $T^4/{\bf Z}_2 $ and of blow-up formulas including 
$A_{N-1}$ theta series with level $N$.
\end{abstract}
\end{center}
\end{titlepage}
\section{Introduction}
\label{sec:1}
\setcounter{equation}{0}

${\cal N}=4$ supersymmetric Yang-Mills theory on $4$ dimensional manifold  
has the largest number of
supersymmetry if it does not couples to gravity.
This theory is one of the well-known examples of the theories 
that are conformally invariant and finite.
In \cite{M-O}, Montonen and Olive conjectured that the theory has a 
kind of duality under the inverse operation of gauge coupling constant 
$\frac{4\pi i}{g^{2}}\rightarrow -\frac{g^{2}}{4\pi i}$. 
In \cite{vafa-witten}, Vafa and Witten considered the twisted version 
of ${\cal N}=4$
supersymmetric Yang-Mills theory and generalized the Montonen-Olive 
duality, by combining the coupling constant with theta angle 
$\tau:=\frac{4\pi i}{g^{2}}+\frac{\theta}{2\pi}$, 
to the symmetry of the partition function under the subgroup 
of $SL(2,{\bf Z})$ ($S$-duality). Using this conjecture as an assumption, 
they derived the partition 
function of the theory on a complex surface with ample canonical bundle.
The key observation of the derivation is the mass perturbation of 
${\cal N}=4$ theory down to ${\cal N}=1$ theory. 
This perturbation does not change the partition function.  
With this operation, the partition function is 
separated into the following two factors 
(the same method was also used in \cite{witten}).
One part is the contribution from bulk 
(non-zero locus of the section of the canonical bundle), or massive part.
This contribution is expressed in terms of a kind of  
G\"ottsche formula \cite{nak, mukai, vafa-witten}, using analogy 
with the result of $K3$ surface (note that $K3$ surface has trivial canonical 
bundle). 
The other part is the contribution from cosmic string
 (zero locus of the section of the canonical bundle), or massless part.
It is determined from the observation of the change of partition 
function under blow-up of complex surface, originally derived in 
\cite{yoshioka}.    
Finally, they combined these two factors to satisfy the desired 
property imposed from the $S$-duality conjecture.
After their work, deeper analyses and further applications
were carried out in \cite{bonelli, dijkgraaf, jin, kap, laba, lozano, qin, 
m-v, sako, yoshi, yoshi2, yoshihecke, yoshisun, prime}.
Especially in \cite{sako}, Sako and T.S. (one of the authors) 
revealed the fact that
Euler number of instanton moduli space and Seiberg-Witten invariants are
connected in the framework of Vafa-Witten theory. 

In our previous work \cite{jin},
we derived the partition function of ${\cal N}=4$ supersymmetric 
Yang-Mills theory
on orbifold-$T^4/{\bf Z}_2 $ for $SU(2)$ gauge group  from the point of view 
of orbifold construction. Our idea is very simple. In orbifold construction, 
we first obtain a quotient space $S_{0}$ which has trivial canonical 
bundle and sixteen double singularities. Next, we blow-up sixteen 
singularities (they turn into sixteen ${\cal O}(-2)$ curves) and 
obtain a Kummer surface, a special class of $K3$ surfaces. 
Therefore we speculated that the $SU(2)$ partition function of $K3$ 
surface should be expressed as the product of the bulk contribution  
of the quotient space $S_0$ and of blow-up formulas coming 
from blowing up the sixteen singularities of $S_{0}$. 
Our speculation was proved to be true in this case \cite{jin}, with the 
aid of some identities between Jacobi's theta functions and Dedekind's
eta function.

In this paper, we try to generalize this construction 
to the case of $SU(N)$ gauge group  ($N$ is odd prime number).
For $SU(N)$ theory it is well-known that
the partition function on $K3$
is given by a Hecke transformation of order $N$ of $1/\eta^{\chi(K3)}(\tau)$.
Analogous results on gauge group $SU(N)/{\bf Z}_{N}$ for $K3, \frac{1}{2}K3 
$ and $T^{4}$ surfaces are also known.
This fact was derived and used from physical side 
in \cite{vafa-witten, m-v, lozano},
and was confirmed mathematically in \cite{yoshihecke}.
From this starting point, we asked whether $K3$ partition function for 
$SU(N)$ (or $SU(N)/{\bf Z}_N $) can be expressed as the  
sum of the product of the following two factors:
the bulk contribution of $T^4/{\bf Z}_2=S_0$ 
and ${\cal O}(-2)$ curve blow-up formula from 16 orbifold singularities.
In this case,  our answer turns out to be negative if we use the 
${\cal O}(-2)$ curve blow-up formulas.
But we found that the above construction is possible if we use 
the ${\cal O}(-N)$ curve blow-up formula 
derived in \cite{kap} instead. Here we mean 
${\cal O}(-N)$ curve blow-up formula 
by the contribution to the partition function from the ${\bf P}^{1}$ 
in the surface with self-intersection number $-N$.
The formula we obtained is given as follows:
\beq
Z^{K3}_k=\sum_{j=0}^{N-1}Z_j^{S_0}Z_{k-j+1}^B,
\label{j1}
\enq
\beq
Z_j^B=\sum_{\{\beta_{l}\}} a_{\{\beta_{l}\}}^{j} 
\prod_{l=1}^{16} \frac{\theta_{(N)}^{\beta_{l}}(\tau)}{\eta^N(\tau)}.
\label{j2}
\enq
Here $Z_j^X$ stands for the partition function on $X$
with 't Hooft flux $v\in H^2(X,{\bf Z}_N),\;\;v^{2}=2j\;\; (mod\; N)$.
$\frac{\theta_{(N)}^{\beta_l}(\tau)}{\eta^N(\tau)} $ 
is the ${\cal O}(-N) $ curve blow-up formula 
labeled by $\beta_l$ \cite{kap}.
We emphasize that $a_{\{\beta_{l}\}}$ is an integer.
The key relation in our derivation of (\ref{j2}) is the well-known 
``denominator identity'' in the theory of affine Lie algebras.
This fact seems to suggest to have a close relation between 
${\cal N}=4$ supersymmetric Yang-Mills theory 
and rational conformal field theory (affine Lie algebra),
as was already pointed out in \cite{vafa-witten}.

\paragraph{${\cal O}(-N)$ curve blow-up}~\\
In this paper, we found the following two key relations.
One is given by,
\beq
\frac{1}{\eta(\frac{\tau}{N})}=\frac{\theta^{2}_{A_{N-1}}(\tau)}{\eta^N(\tau)},
\label{jin3}
\enq
where $\theta^{2}_{A_{N-1}}(\tau)$ is a theta series related to affine Lie 
algebra.
The other is the following:
\beq
\theta_{A_{N-1}}(\tau)=\sum_\beta a_{\beta} \theta_{(N)}^\beta (\tau),
\label{jin4}
\enq
where $\theta_{(N)}^\beta(\tau)$ is Kapranov's theta function labeled by 
$\beta$ associated with ${\cal O}(-N)$ curve blow-up \cite{kap}.
These blow-up formulas  correspond to 
the generating functions of the Euler number of 
``relative moduli space'' of ${\bf P}^{1}$ with self intersection number
$-N$ in complex surface. 
By the word ``relative'' we mean to consider 
the configuration localized near ${\bf P}^{1}$ like the case of 
``local mirror symmetry''. 
Here, we again emphasize that the coefficient $a_{\beta}$ is an integer.
Using these two key relations,
we can rewrite $1/\eta(\frac{\tau}{N})$ 
by ${\cal O}(-N) $ curve blow-up formulas and obtain (\ref{j2}).
We cannot still figure out the reason why we face 
${\cal O}(-N) $ curve blow-up formula 
instead of ${\cal O}(-2) $ curve
blow-up formula.  Hence the geometrical interpretation 
of our result should be pursued further.
The fact that $a_{\beta}$ is an integer seems to suggest 
the possibility of geometrical interpretation.
We can easily see  
that ${\cal O}(-N)$ curve blow-up formula appearing in our formulas
comes from level $N$ representation of affine $SU(N)$ characters.
On the other hand, level $k$ characters of affine $SU(N)$ algebra
appears in ${\cal N}=4$ super Yang-Mills theory with gauge group $U(k)$ 
on $A_{N-1}$ ALE space \cite{vafa-witten, naka}.
From this point of view, 
the appearance of ${\cal O}(-N)$ curve blow-up formula 
seems to be natural, because we consider $SU(N)$ gauge group 
\cite{vafa-witten, naka}.

This paper is organized  as follows.
In Sec.2, we review our previous work on $SU(2)$ gauge group  
and derive the bulk contribution of $S_0$ for gauge group 
$SU(N)/{\bf Z}_{N}$.
In Sec.3, we introduce the key identity (\ref{jin3}) and prove  
it by using a denominator identity
of affine Lie algebra.
In Sec.4, we give a key conjecture and rewrite $1/\eta(\frac{\tau}{N})$
in terms of ${\cal O}(-N)$ curve blow-up formulas.
In Sec.5, we derive the partition function of $K3$ for $SU(N)/{\bf Z}_N$
by using the results of Sec.3, 4.
In Sec.6, we conclude and discuss remaining problems.

\section{Review of the $SU(2)$ case and Contribution from Untwisted Sector}
\label{sec:2}
\setcounter{equation}{0}

In this section, we briefly review the geometrical background of
orbifold $T^4/{\bf Z}_2 $ \cite{Asp,Fukaya} and derive the 
contribution from the  untwisted sector of
${\cal N}=4$ Supersymmetric Yang-Mills Theory
on orbifold-$T^4/{\bf Z}_2$ for $SU(N)$ \cite{vafa-witten,
dijkgraaf,lozano,yoshi2,qin}.
Next, we review our previous work on the reconstruction of 
 $K3$ partition function for $SU(2)$ case,
where we use ${\cal O}(-2) $ curve blow-up formula \cite{yoshioka, jin}.

\subsection{Orbifold $T^4/{\bf Z}_2 $}

In this subsection, we construct $K3$ surface from $T^4$ surface by 
orbifold construction, in keeping canonical bundle 
$K_X=0 $ trivial \cite{Asp,Fukaya}.

First, we consider $T^4$ surface. $T^4$ surface obviously 
has a trivial canonical bundle
$K_{T^4}=0$, but has nontrivial Picard groups coming 
from $dim(H^{1}(T^{4},{\bf C}))=4$.
After the ${\bf Z}_2 $-identification, we obtain a quotient 
space  $T^4/{\bf Z}_2=:S_0 $,
which has also a trivial canonical bundle and trivial Picard groups, 
because  $H^{1}(T^{4},{\bf C})$ is not invariant under the action 
of ${\bf Z}_{2}$.
Note that $S_0 $ has sixteen orbifold singularities.
Therefore, we have to resolve these singularities.
For this purpose, we blow-up these singularity points and obtain 
sixteen  ${\cal O}(-2) $ curves (we call this process ${\cal O}(-2) $ 
curve blow-up).
After this process, $S_{0}$ turns into a Kummer surface, a special class of 
compact and smooth $K3$ surfaces.

Here, we write down the change of Euler numbers of surfaces under the 
process we have just described, 
\begin{equation}
\chi(T^4)=0 \to \chi(S_0)=8 \to \chi(K3)=24.
\end{equation}

\subsection{Contribution from Untwisted Sector}

\paragraph{General Structure of Vafa-Witten Conjecture}
~\\
Following \cite{vafa-witten,lozano,qin,yoshi}, we review the
general structure of Vafa-Witten conjecture.
For twisted ${\cal N}=4$ $SU(N)/{\bf Z}_N$ gauge 
theory with 't Hooft flux $v\in H^{2}(X,{\bf Z}_{N})$ on $X$,
Vafa and Witten showed that the partition function of this theory is 
given by the formula: 
\begin{equation}
Z^X_v(\tau):= q^{-{\frac{N\chi(X)}{24}}}\sum_k \chi({\cal N}(v,k))q^k
\;\;\;(q:=\exp(2\pi i \tau)),
\end{equation}
where ${\cal N}(v,k)$ is 
the moduli space of anti-self-dual connections
associated to $SU(N)/{\bf Z}_N$-principal bundle with
{}'t Hooft flux $v$ 
and fractional instanton number $k\in Z/2N$.
$\tau$ is the gauge coupling constant including theta angle and 
$\chi(X)$ is Euler number of $X$.
With this result, they conjectured the behavior 
of the partition functions under the action of $SL(2,{\bf Z})$ on 
$\tau$. The most important formula of this conjecture is given by,  
\begin{equation}
Z^X_v\left(-\frac{1}{\tau}\right)=N^{-\frac{b_2(X)}{2}}
\left(
\frac{\tau}{i}
\right)^{-\frac{\chi(X)}{2}}
\cdot
\sum_{u\in H^2(X,Z_N)}
\zeta_N^{u\cdot v}Z^X_u(\tau),
\label{vw}
\end{equation}
where $\zeta_N=\exp(\frac{2\pi i}{N}) $.

For later use, we introduce the notation:
\[
Z_{SU(N)}^X(\tau):= \frac{1}{N} Z_0^X(\tau),
\]
\begin{equation}
Z_{SU(N)/{\bf Z}_N}^X(\tau):= \sum_{u\in H^2(X,Z_N)} Z_u^X(\tau).
\end{equation}
In this notation, we can obtain the following 
formula from (\ref{vw}):
\begin{equation}
Z^X_0\left(-\frac{1}{\tau}\right)=N^{-\frac{b_2(X)}{2}}
\left(
\frac{\tau}{i}
\right)^{-\frac{\chi(X)}{2}}
Z_{SU(N)/{\bf Z}_N}^X(\tau).
\label{so3}
\end{equation}
This formula is one of the key points on their explicit determination 
of the form of the partition function of complex surface with ample 
canonical bundle. 

\paragraph{Partition Function of the Untwisted Sector of $S_0$}
~\\
Next,
we derive the partition function of the theory on the intermediate 
quotient space $S_0$ as a first step of generalization of 
our previous work.
At first, we consider the moduli space ${\cal N}(v,k)$ of 
$X=T^4 $.
In this case, we can identify the moduli space ${\cal N}(v,k)$
with the moduli space $\bar{\cal M}_H(N,c_1,c_2) $
of rank $N$ stable sheaves $E$
with Chern classes $c_1,c_2\;\;(c_1=v~{\rm mod}~N $ and 
$k=c_2-\frac{(N-1)v^2}{2N}) $ \cite{yoshi,yoshi2}.
At this point, we need to note that
we restrict $N$ to prime numbers throughout 
this paper. This condition makes the structure of $\bar{\cal M}_H(N,c_1,c_2) $
tractable.

According to \cite{yoshi2},
the moduli space of rank $N\; (N: \mbox{prime})$ stable sheaves $E$ of $V$
is given as follows. First, We introduce Mukai vector in 
$\oplus{}_j H^{2j}(X,{\bf Z})$, 
\begin{equation}
V=ch(E)\sqrt{td_X}=N+c_1+\frac{c_1^2-2c_2}{2}.
\end{equation}
We remark here that in our case $X=T^4$, $td_X=1$.
Then we introduce the inner product of Mukai vector, 
\begin{eqnarray}
<V^2>&=&-\int_X
\left(N+c_1+\frac{c_1^2-2c_2}{2}\right)\vee
\left(N+c_1+\frac{c_1^2-2c_2}{2}\right)
\nonum
\\
&=&2Nc_2-(N-1)c_1^2.
\end{eqnarray}
Here we use a symmetric bilinear form on 
$\oplus{}_j H^{2j}(X,{\bf Z}) $:
\begin{eqnarray}
<x,y>&=&-\int_X(x\vee y)
\nonum
\\
&=&
  \int_X(x_1y_1-x_0y_2-x_2y_0),
\end{eqnarray}
where $x=x_0+x_1+x_2,x_j\in H^{2j}(X,{\bf Z}) $
and $x\vee =x_0-x_1+x_2$.

With these preparation, we can describe the moduli space 
$\bar{\cal M}_H^X(c_{1},c_{2})$ explicitly in terms of Hilbert 
scheme of $n$ points $X^{[n]}$ of $X$,
\begin{equation}
\bar{\cal M}_H^X(c_{1},c_{2})\cong {\hat X}\times (X)^{[\frac{<V^2>}{2}]}
= {\hat X}\times(X)^{[Nc_2-\frac{(N-1)c_1^2}{2}]},
\end{equation}
where $H$ is polarization of $X$, and
${\hat X}$ is the $T^{4}$ surface dual to $X$, which represents 
non-trivial Picard group  
\cite{yoshi2}. 
Here, we have to notice that $\bar{\cal M}_H^{K3}(c_{1},c_{2})$
is also diffeomorphic to $(K3)^{[\frac{<V^2>}{2}+1]}$.

Since $S_{0}$ has the trivial canonical bundle like $K3$ and $T^{4}$, it 
doesn't have cosmic strings, which are given by zero locus of the section of 
the canonical bundle. From the point of view of \cite{vafa-witten}, this
fact leads us to expect that the corresponding moduli space of $S_{0}$   
is also diffeomorphic to Hilbert scheme of points on $S_{0}$,
\begin{equation}
\bar{\cal M}_H(c_{1},c_{2})\cong (S_0)^{[\frac{<V^2>}{2}]}
= (S_0)^{[Nn-\frac{(N-1)v^2}{2}]},
\end{equation}      
where we used the fact that the Picard group of $S_{0}$ is trivial
\footnote[1] 
{Precisely speaking, we have to take care of non-trivial Todd class 
of $S_{0}$. But we don't know the precise definition of Todd class 
of $S_{0}$. Moreover, this correction does not affect the results of 
computation severely. Therefore, we neglect here this correction.}.
Using this assumption and G\"ottsche formula, 
we can evaluate the partition function of
$S_0$, i.e., the contribution from the untwisted sector 
with  $v^2=2j$ (mod $N$) type,
\begin{eqnarray}
Z^{S_{0}}_{j}(\tau)
&=&
q^{-\frac{1}{3N}}\sum_{v^2\equiv j(mod~N),n} e(\bar{\cal M}_H(V))
q^{n-\frac{N-1}{N}j}  
\nonum
\\
&=&
q^{-\frac{1}{3N}}\sum_n e((S_0)^{[Nn-(N-1)j]})
q^{n-\frac{N-1}{N}j}
\nonum  
\\
&=&
q^{-\frac{1}{3N}}\sum_m e((S_0)^{[m]})
\frac{(q^{\frac{1}{N}})^m+\zeta_N^{-j}(\zeta_Nq^{\frac{1}{N}})^m+\cdots
+\zeta_N^{-j(N-1)}(\zeta_N^{N-1}q^{\frac{1}{N}})^m
}{N}
\nonum
\\
&=&
\frac{1}{N}\left(\frac{1}{\eta^8(\frac{\tau}{N})}
+\zeta_N^{\frac{1}{3}}\zeta_N^{-j}\frac{1}{\eta^8(\frac{\tau+1}{N})}
+\cdots+\zeta_N^{\frac{N-1}{3}}\zeta_N^{-j(N-1)}\frac{1}{\eta^8(\frac{\tau+N-1}{N})}\right).
\nonum 
\\  \label{zse}
\end{eqnarray}

For trivial type $v=0$, which corresponds to $SU(N)$ gauge group, 
we follow the discussion in \cite{vafa-witten} and set,
\begin{equation}
Z^{S_{0}}_t(\tau)=C\frac{1}{\eta^8(N\tau)}+\frac{1}{N}
\frac{1}{\eta^8(\frac{\tau}{N})}
+\frac{1}{N}\zeta_N^{\frac{1}{3}}\frac{1}{\eta^8(\frac{\tau+1}{N})}
+\cdots+\frac{1}{N}\zeta_N^{\frac{N-1}{3}}\frac{1}{\eta^8(\frac{\tau+N-1}{N})},
\end{equation}
where $C$ is some unknown constant. 
Note that the term $\frac{1}{\eta^8(N\tau)}$ is obtained from  
$\frac{1}{\eta^8(\frac{\tau}{N})}$ by using the modular transformation 
$\tau \rightarrow -\frac{1}{\tau}$, and should be included to 
satisfy the conjecture (\ref{so3}).

\subsection{${\cal O}(-2) $ Curve Blow-up Formula and $K3 $ Partition Function:
Review of the $SU(2) $ case}
In this subsection, we review the discussion of our previous work on  
$SU(2)$ case \cite{jin}.
Especially, we reconstruct the $K3$ partition function by 
combining the contribution from the untwisted sector (i.e., $S_{0}$) 
for the $SU(2)$ case
with the contribution of blowing up sixteen double singularities of $S_{0}$.
In \cite{jin}, we introduced ${\cal O}(-2) $
curve blow-up formulas as the contribution from the twisted sector:
\begin{equation}
\frac{\theta_{2}(\tau)}{\eta(\tau)^{2}},\;\; 
\frac{\theta_{3}(\tau)}{\eta(\tau)^{2}},\;\;
\frac{\theta_{4}(\tau)}{\eta(\tau)^{2}}.   
\end{equation}
They describe the contribution to the partition function coming from 
${\bf P}^{1}$ with self intersection number $-2$. This ${\bf P}^{1}$ 
appear as the result of blowing up the double singularity in the complex 
surface. To be precise, we rearrange better the above formulas in the 
following way,
\begin{eqnarray}
&&\frac{1}{2}\frac{\theta_{2}(\tau)}{\eta(\tau)^{2}}
=\frac{1}{2}(\sum_{n\in{\bf Z}}q^{\frac{1}{2}(n-\frac{1}{2})^{2}})/
\eta(\tau)^{2}=(\sum_{n\in{\bf Z}}q^{2(n-\frac{1}{4})^{2}})/
\eta(\tau)^{2},
\no\\ 
&&\frac{1}{2}\frac{\theta_{3}(\tau)+\theta_{4}(\tau)}{\eta(\tau)^{2}}
=(\sum_{n\in{\bf Z}}q^{{2}n^{2}})/
\eta(\tau)^{2},\no\\ 
&&\frac{1}{2}\frac{\theta_{3}(\tau)-\theta_{4}(\tau)}{\eta(\tau)^{2}}
=(\sum_{n\in{\bf Z}}q^{{2}(n-\frac{1}{2})^{2}})/
\eta(\tau)^{2}.  
\label{-2}
\end{eqnarray}
Heuristically, these blow-up formulas correspond to the 
contributions from the rank 2 vector bundles 
localized near ${\bf P}^{1}$: ${\cal O}(nE)\oplus{\cal O}((-n+\frac{1}{2})E),
\;\;{\cal O}(nE)\oplus{\cal O}(-nE),\;\;{\cal O}(nE)\oplus{\cal O}((-n+1)E)$ 
respectively. Here, we denote by $E\;(E\cdot E=-2)$, the divisor 
corresponding to the above ${\bf P}^{1}$. Notice that the second Chern 
classes  of these vector bundles are given by $2n(n-1/2),\;\;2n^{2},\;\;
2n(n-1)$ respectively. These numbers coincide with the powers of $q$ in 
theta series except for constants in $n$. We will discuss such 
correspondence more generally in the next section. 
The factor $\frac{1}{\eta(\tau)^{2}}$ comes from the boundary part 
of the moduli space of rank 2 stable sheaves localized near ${\bf P}^{1}$.
The rank two stable sheaves $E_{b}$ on the boundary of the moduli space are 
written as direct sum of two ideal sheaves 
$I_{Z_{1}}, I_{Z_{2}}$ of Hilbert scheme 
of points on the normal bundle $N({\bf P}^{1})$ of the ${\bf P}^{1}$.
Since the total space of $N({\bf P}^{1})$ has only $1$ 2-cycle 
${\bf P}^{1}$ as the nontrivial cycles (0-cycle is already counted 
in the contribution from $S_{0}$), it is obvious from G\"ottsche formula 
that the corresponding contribution is given by 
$\frac{1}{\eta(\tau)^{2}}$.
For mathematically rigorous discussion, see \cite{yoshioka, qin}. 

Now we separate 16 double singularities into two groups that consist of 
8 double singularities. Then we blow-up 8 singularities in each group 
at a time. The resulting contributions of these operations are given by the 
following formulas:   
\begin{eqnarray}
{\tilde Z}_1(\tau)
&=&
\frac{\theta_2^8(\tau)(\theta_3^8(\tau)+\theta_4^8(\tau))}{2^{9}
\eta^{32}(\tau)},
\\
{\tilde Z}_2(\tau)
&=&
\frac{\theta_2^8(\tau)(\theta_3^8(\tau)-\theta_4^8(\tau))}{2^{9}
\eta^{32}(\tau)},
\\
{\tilde Z}_3(\tau)
&=&
\frac{\theta_3^8(\tau)\theta_4^8(\tau)}{\eta^{32}(\tau)}.
\label{8}
\end{eqnarray}
In the derivation of ${\tilde Z}_3(\tau)$, we used the Vafa-Witten conjecture
on modular property of the partition function. 
Using these formulas, we reconstructed $K3$ partition function: 
\begin{eqnarray}
&&Z^{K{3}}_{even}(\tau)=
4\left(
Z^{S_{0}}_{odd}(\tau){\tilde Z}_2(\tau)
+
Z^{S_{0}}_{even}(\tau){\tilde Z}_1(\tau)
\right),\no\\
&&Z^{K3}_{odd}(\tau)=
4\left(
Z^{S_{0}}_{odd}(\tau){\tilde Z}_1(\tau)
+
Z^{S_{0}}_{even}(\tau){\tilde Z}_2(\tau)
\right).
\label{evev}
\end{eqnarray} 
\begin{equation}
Z^{K3}_t(\tau)=
\frac{1}{4}\frac{1}{\eta^8(2\tau)}{\tilde Z}_3(\tau)
+4\left(
Z^{S_{0}}_{even}(\tau){\tilde Z}_2(\tau)
+
Z^{S_{0}}_{odd}(\tau){\tilde Z}_1(\tau)
\right).
\label{0}
\end{equation}
To compare these results with those of Vafa-Witten,
we rewrite these partition functions using identities
between Dedekind's eta function and Jacobi's theta functions:
\begin{equation}
\frac{\theta^{8}_{3}(\tau)\theta_{4}^{8}(\tau)}{\eta^{32}(\tau)}=
\frac{1}{\eta^{16}(2\tau)},\;\;\; 
\frac{\theta^{8}_{2}(\tau)\theta_{3}^{8}(\tau)}{2^{8}\eta^{32}(\tau)}=
\frac{1}{\eta^{16}(\frac{\tau}{2})},\;\;\;
\frac{\theta^{8}_{2}(\tau)\theta_{4}^{8}(\tau)}{2^{8}\eta^{32}(\tau)}=
e^{\frac{8\pi i}{3}}\frac{1}{\eta^{16}(\frac{\tau+1}{2})}.
\label{et2}
\end{equation}
Then we obtain the well-known results in \cite{vafa-witten} as expected,
\begin{eqnarray}
&&Z^{K3}_t(\tau)=
\frac{1}{4}\frac{1}{\eta^{24}(2\tau)}
+\half\frac{1}{\eta^{24}(\frac{\tau}{2})}
+\half\frac{1}{\eta^{24}(\frac{\tau}{2}+\half)},\no\\
&&Z^{K3}_{even}(\tau)=
\half\frac{1}{\eta^{24}(\frac{\tau}{2})}
+\half\frac{1}{\eta^{24}(\frac{\tau}{2}+\half)},\no\\
&&
Z^{K3}_{odd}(\tau)=
\half\frac{1}{\eta^{24}(\frac{\tau}{2})}
-\half\frac{1}{\eta^{24}(\frac{\tau}{2}+\half)}.
\end{eqnarray}

\section{From Bulk to Blow-up}
\label{sec:3}
\setcounter{equation}{0}

In the rest of this paper, we generalize our previous work 
reviewed in Sec.2 to the $SU(N)$ case.
Especially in the following two sections,
we introduce the blow-up formula
that appears in our theory.
In this section, we first derive the blow-up formula in the 
simplest case, and then introduce key identities which motivated  
us to generalize the identities in (\ref{et2}).  
Finally, we give a proof of these identities with  
the aid of the denominator identity of affine Lie algebra \cite{mac, kac}.    

\subsection{${\cal O}(-d)$ blow-up formula for the ${\cal N}=4$ $SU(N)$ gauge 
theory} 
In this subsection, we sketch the derivation of ${\cal O}(-d)$ blow-up 
formula for the ${\cal N}=4$ $SU(N)$ gauge theory. Roughly speaking, 
the blow-up formula is the generating function of Euler number of the 
moduli space of the rank $N$ stable sheaves localized near ${\bf P}^{1}$ 
in a complex surface with self-intersection number $-d$.   
Heuristically, the bulk part of the moduli space 
is separated into connected components labeled by the following 
rank $N$ vector bundle on a complex surface,
\begin{eqnarray}
{\cal O}(m_{1}E_{d})\oplus{\cal O}(m_{2}E_{d})\oplus
\cdots\oplus{\cal O}(m_{N}E_{d}),\no\\
m_{i}\in{\bf Z},\;\;m_{1}+m_{2}+\cdots+m_{N}=0,
\label{bulkN}
\end{eqnarray}  
where $E_{d}\;\; (E_{d}\cdot E_{d}=-d)$ stands for the divisor corresponding 
to the above ${\bf P}^{1}$. The condition $\;\;m_{1}+m_{2}+\cdots+m_{N}=0$
is imposed by the vanishing first Chern class. The second Chern class of
the bundle in (\ref{bulkN}) is given by,
\begin{equation} 
(\sum_{i<j}m_{i}m_{j})E_{d}\cdot E_{d}=\frac{d}{2}\sum_{i=1}^{N}(m_{i})^{2},
\label{2c}
\end{equation}
where we used the condition $m_{1}+m_{2}+\cdots+m_{N}=0$. 
But the lattice 
$\{(m_{1},m_{2},\cdots,m_{N})\in{\bf Z}^{N}|\;\;m_{1}+m_{2}+\cdots+m_{N}=0\}$
with usual Euclid metric in ${\bf R}^{N}$ is nothing but the $A_{N-1}$ 
root lattice. Therefore, the connected components in (\ref{bulkN}) are 
labeled by a root vector $n_{1}\alpha_{1}+\cdots+n_{N-1}\alpha_{N-1}$. 
The second Chern class in (\ref{2c}) turns out to be  
$\frac{d}{2}({}^{t}n A_{N-1} n)$. 
Here, we denote by $A_{N-1}$ Cartan matrix of $A_{N-1}$ 
root lattice. It can be shown that Euler number of each connected 
component is equal to $1$ \cite{kap}. Then it follows 
that the contribution from  
the bulk part of the moduli space is given by the theta series:
\begin{equation}
\theta_{A_{N-1}}^{0,d}(\tau):=\sum_{m\in {\bf Z}^{N-1}}
q^{\frac{d}{2}({}^{t}m A_{N-1}m)}.
\label{d0}   
\end{equation}
Next, we consider the contribution from the boundary part of 
the moduli space. As we have discussed in the previous section, the rank 
$N$ stable sheaves in the boundary parts of the moduli space are described  
by direct sum of $N$ ideal sheaves $I_{Z_{1}}, I_{Z_{2}}, \cdots ,
I_{Z_{N}}$ of Hilbert scheme of points on $N({\bf P}^{1})$. Therefore,
we can conclude that the contribution to the generating function is given by 
$\frac{1}{\eta(\tau)^{N}}$, using the same discussion as in the $SU(2)$
case. Combining the two contributions, we are led to the ${\cal O}(-d)$ 
blow-up formula for ${\cal N}=4$ $SU(N)$ gauge theory:
\begin{equation}
\frac{\theta_{A_{N-1}}^{0,d}(\tau)}{\eta(\tau)^{N}}
=\sum_{m\in {\bf Z}^{N-1}}
q^{\frac{d}{2}({}^{t}m A_{N-1}m)}/
(q^{\frac{N}{24}}\prod_{n=1}^{\infty}(1-q^{n})^{N}).
\label{dblow}
\end{equation}
Now, we are at the most important point of generalization of our previous 
work on ${\cal N}=4$ $SU(2)$ gauge theory. Using the same method 
in the previous section, it can be easily seen that the partition function of 
${\cal N}=4$ $SU(N)/{\bf Z}_{N}$ gauge theory on $K3$ surface consists of 
the functions $1/\eta(\frac{\tau+j}{N})\;\;(j=0,1,\cdots,N-1)$. 
Therefore, we have to 
search for some identities that relate $1/\eta(\frac{\tau+j}{N})$ 
to $\Theta_{A_{N-1}}(\tau)/\eta(\tau)^{N}$ (we denote by 
$\Theta_{A_{N-1}}(\tau)$ some variant of $\theta_{A_{N-1}}^{0,d}(\tau)$).
Fortunately, we found the desired identities. Let us discuss this point 
in the following subsections.    
  
\subsection{Key Identity}
In this subsection, we introduce the identities that represent 
the functions ${1}/{\eta(N\tau)}$ and $1/{\eta(\frac{\tau}{N})}$
in terms of the function ${\Theta_{A_{N-1}}(\tau)}/{\eta^{N}(\tau)}$. 
In the $SU(2)$ case, this identity is simply given by,
\begin{equation}
\frac{1}{\eta(2\tau)}=\frac{\theta_4(2\tau)}{\eta^2(\tau)},
\label{2th1}
\end{equation}
or
\begin{equation}
\frac{1}{\eta(\frac{\tau}{2})}=
\half\frac{\theta_2(\frac{\tau}{2})}{\eta^2(\tau)}.
\label{2th2}
\end{equation}
Note that the right hand sides of these identities have the form 
$\Theta_{A_{1}}(\tau)/\eta^2(\tau) $. 
Then we searched for a possible 
generalization of these identities to the case of 
$SU(N)$ gauge group. 
After some trial and error, we found the following 
key identities:
\begin{equation}
\frac{1}{\eta(N\tau)}=\frac{\theta_{A_{N-1}}^1(\tau)}{\eta^N(\tau)},
\label{Nth1}
\end{equation}
or alternatively
\begin{equation}
\frac{1}{\eta(\frac{\tau}{N})}=\frac{\theta_{A_{N-1}}^2(\tau)}{\eta^N(\tau)},
\label{Nth2}
\end{equation}
where we introduced $A_{N-1}$ theta functions,
\begin{equation}
\theta_{A_{N-1}}^0(\tau):= \sum_{m\in{\bf Z}^{N-1}}
q^{\frac{1}{2}{}^tm A_{N-1} m},
\label{Ath0}
\end{equation}
\begin{equation}
\theta_{A_{N-1}}^1(\tau):= \sum_{m\in{\bf Z}^{N-1}}
q^{\frac{1}{2}{}^tm A_{N-1} m}
e^{2\pi i m\cdot \delta},
\label{Ath1}
\end{equation}
\begin{equation}
\theta_{A_{N-1}}^2(\tau):= \sum_{m\in{\bf Z}^{N-1}}
q^{\frac{1}{2}{}^t(m+\frac{\rho}{N} )A_{N-1} (m+\frac{\rho}{N})}.
\label{Ath2}
\end{equation}
Here, we introduced the vector $\rho$ in ${\bf Z}^{N-1}\otimes_{\bf Z}{\bf Q}$:
\begin{equation}
\delta:= \frac{1}{N}(1,\ldots,1),~
\rho=N A_{N-1}^{-1} \delta,
\end{equation}
which is the same as the usual $\rho$ given  
by a half of the sum of the positive roots of Lie algebra $A_{N-1}$. 
Note that we can obtain (\ref{Nth1}) by applying  
the modular transformation
$\tau \rightarrow -\frac{1}{\tau}$ to (\ref{Nth2}).

\subsection{Denominator (Macdonald) Identity and Affine Lie Algebra}

In this subsection, we give the proof of the identity in (\ref{Nth2}).
As a warming-up, we consider the $SU(2)$ case given in (\ref{2th2}).
In this case, it is well-known that $\theta_2(\frac{\tau}{2}) $ 
has the following product formula
(which follows from Jacobi's triple product identity):
\beq
\theta_2(\frac{\tau}{2})=2q^{\frac{1}{16}}
\prod_{n=1}^\infty (1-q^{\frac{n}{2}})(1+q^{\frac{n}{2}})^2
=2\frac{\eta^2(\tau)}{\eta(\frac{\tau}{2})}.
\enq
Hence the proof of (\ref{2th2}) is completed.

Next, we turn into the $SU(N)$ case given in (\ref{Nth2}). 
In this case, We already have the celebrated denominator identity of 
affine Lie algebra, that corresponds to
higher rank version of Jacobi's triple product identity \cite{kac}:
\begin{thm}
{\bf (Denominator Identity)}
\beq
\prod_{\alpha\in \Delta_+}(1-e(-\alpha))^{{\rm mult}~ \alpha}
=\sum_{w\in W}\epsilon(w)e(w(\rho)-\rho).
\label{deno}
\enq
\end{thm}
For convenience of readers, we introduce here the notations of the 
affine Lie algebras.
\begin{notn}
Notations \cite{kac, mac},\\
$\Delta:$ set of roots,~
$\Delta_+\subset \Delta:$ set of positive roots,~
$l:$ rank of Cartan matrix\\
mult $\alpha:$ multiplicity of $\alpha \in \Delta_+$,\\
$W:$ Weyl Group,~
$w\in W:$ its element,~
$\epsilon(w)=(-1)^{l(w)},$~
$l(w):$ length of $w$,\\
$\rho:$ half of the sum of the positive roots,\\
$h^\vee$: dual Coxeter number. 
\end{notn}
From now on, the symbol ``prime (${}^\prime$)'' denotes restriction
to classical Lie algebra associated with affine Lie algebra.

Since Weyl group $W$ of affine Lie algebra is given by a 
semi-direct product of 
classical Weyl group $W^{\prime}$ and of classical root lattice $L$, 
we can rewrite (\ref{deno}) into the following form:
\begin{cor}
\beq
q^{\frac{\vert \rho\vert^2}{2h^\vee}}
\prod_{n\ge 1}((1-q^n)^{l}
\prod_{\alpha\in \Delta^\prime}(1-q^n e(\alpha)))
=\sum_{\alpha\in L}\chi^\prime(h^\vee \alpha)
q^{\frac{\vert\rho+h^\vee\alpha \vert^2}{2h^\vee}},
\label{hrv}
\enq
where
\beq
\chi^\prime(\lambda)=
\frac{\sum_{w\in W^\prime}\epsilon(w)e(w(\lambda+\rho)-\rho)}
{\prod_{\alpha\in \Delta^\prime_+}(1-e(-\alpha))}.
\enq
\end{cor}
Note that $q$ stands for $e(\alpha_{0})$ ($\alpha_{0}$ is 
the null root of affine root system) 
and that $\vert u\vert^2 $ is the square length of a vector $u$ in $L$. 
In our case, $\vert\sum_{j=1}^{N-1} m_j \alpha_j \vert^2= {}^tmA_{N-1} m$.\\
With these settings, we can  prove the identity in (\ref{Nth2}).\\
{\bf Proof of (\ref{Nth2})}\\
Let us apply the formula (\ref{hrv}) to the affine $A_{N-1}$
Lie algebra. In this case, we have
\beq
h^\vee=N,\;\;l=N-1,\;\;\frac{\vert \rho\vert^2}{2h^\vee}
=\frac{N^2-1}{24}.
\enq
Next, we specialize the variable $e(\alpha)$ as follows:
\beq
e(\alpha)=\exp(\frac{2\pi i\alpha\cdot \rho}{N}).
\enq
Under this specialization, $\chi^\prime(N\alpha)$ turns into $1$ as a
consequence of the denominator identity applied to the classical Lie algebra 
$A_{N-1}$.
Therefore, the identity in Corollary 1 turns into an identity between 
the following two $q$-series:
\begin{eqnarray}
&&
q^{\frac{N^2-1}{24}}\prod_{n=1}^{\infty}(1-q^n)^{N-1}\prod_{j=1}^{N-1}
(1-q^n\zeta^j_N)^N
\nonum
\\
&&=
\sum_{m\in (N{\bf Z})^{N-1}}q^{\frac{1}{2N}{}^t(m+\rho)A_{N-1}(m+\rho)}.
\end{eqnarray}
Substituting $q^{\frac{1}{N}}$ into $q$, we obtain the desired 
formula (\ref{Nth2}). $\Box$
\\
Note that this proof was first given by Macdonald (p.120,121) \cite{mac}.

We have thus shown that the function 
$1/\eta(\frac{\tau}{N})$ can be described by 
$ \theta_{A_{N-1}}^{2}(\tau)/\eta^N(\tau)$ using the key identity.
But this identity is not enough to rewrite
the function $1/\eta(\frac{\tau}{N}) $ 
in terms of the blow-up formula, because $\theta_{A_{N-1}}^{2}(\tau)$ contains 
all the integer powers of $q^{\frac{1}{N}}$. 
In the next section, we complete our attempt
to rewrite $1/\eta(\frac{\tau}{N})$ 
in terms of the blow-up formula of $SU(N)/{\bf Z}_{N}$ gauge theory.  
However, we have shown 
that this theory is closely related to the affine Lie algebras,
as was already pointed out in \cite{vafa-witten}.

\section{${\cal O}(-N) $ Curve Blow-up}
\label{sec:4}
\setcounter{equation}{0}

We continue the discussion on the blow-up formula of $SU(N)/{\bf Z}_{N}$ 
gauge theory.
Especially in this section, we try to rewrite
the theta function $\theta_{A_{N-1}}^{2}(\tau)$ in the previous section
as the sum of the ${\cal O}(-N)$ curve blow-up formulas of $SU(N)/{\bf Z}_{N}$ 
gauge theory \cite{jin,kap}.
First, we introduce the key conjecture which we obtained after some trial 
and error.
Next, we show some explicit examples of this conjecture.  

\subsection{Key Conjecture}
In the previous section, we obtained  
an identity that relates  $1/\eta(\frac{\tau}{N})$ to the similar function 
in (\ref{dblow}).
However in the geometrical context,
theta function $\theta^2_{A_{N-1}}(\tau) $ is not the expected theta function
appearing in the blow-up formula.
In the $N=2$ case, we fortunately have the following duplication formula:
\beq
\theta^2_{A_1}(\tau)^2=\half\theta_2(\tau)\theta_3(\tau).
\label{dup}
\enq 
That is why we have introduced (\ref{-2}) directly instead of 
$\frac{\theta^2_{A_1}(\tau)}{\eta^2(\tau)} $.
As we have already mentioned, relations in
(\ref{-2}) are ${\cal O}(-2) $ curve blow-up 
formulas of $SU(2)/{\bf Z_{2}}$ gauge theory and  
they are  consistent with geometrical context (recall Sec.2.1).
In this principle, we first tried to find similar 
identities for general $N$ case
as (\ref{dup}), 
where we can rewrite $\theta^2_{A_{N-1}}(\tau) $ 
in terms of  
the theta function appearing in ${\cal O}(-2) $ curve blow-up formula.
For this purpose, we refer here the theorem of Kapranov 
on the general form of ${\cal O}(-d)$ blow-up formulas of 
 $SU(N)/{\bf Z}_{N}$ gauge theory \cite{kap}:\\
\begin{thm}
[{\bf Kapranov}]
${\cal O}(-d)$-curve blow-up formula for $SU(N)/{\bf Z}_{N}$ gauge theory is 
given by the $A_{N-1}$ theta series with level $d$: 
\beq
\sum_{a\in L}
q^{\Psi(a,f)-d\Psi(a,a)/2},
\label{kap}
\enq 
where $L$ is the $A_{N-1}$ root lattice, $f$ is an element of the weight 
lattice and $\Psi(a,b)=-{}^{t}a A_{N-1}b$. 
\end{thm}
\begin{rem}
In Theorem 2, the factor coming from boundaries of 
the moduli space is neglected because Kapranov treated uncompactified 
case in \cite{kap}. Compactified version of  
${\cal O}(-1)$ curve blow-up formula for $SU(N)/{\bf Z}_{N}$ gauge theory 
was first derived by Yoshioka \cite{yoshisun}.
\end{rem}
\begin{rem}
The original version of Theorem 2 in \cite{kap} takes $f$ as an element 
of the root lattice. But in \cite{vafa-witten} where 
${\cal O}(-1)$-curve blow-up 
formula for $SU(2)/{\bf Z}_{2}$ gauge theory is considered, Vafa and Witten 
introduced the two types of blow-up formula:
\begin{equation}
\Theta_{0}:=\sum_{n\in{\bf Z}}q^{n^{2}},\;\;
\Theta_{1}:=\sum_{n\in{\bf Z}}q^{(n-\frac{1}{2})^{2}}.
\end{equation} 
$\Theta_{0}$ and $\Theta_{1}$ correspond to taking $f$ as $0$ and  
$f$ as $\frac{1}{2}\alpha$ respectively. Therefore, it is natural to take 
$f$ as an element of the weight lattice. This generalization is also 
compatible with our treatment of {\cal O}(-2)-curve blow-up 
formula for $SU(2)/{\bf Z}_{2}$ gauge theory.
\end{rem}

Then we applied Theorem 2 to the $d=2$ case 
and tried to find a formula which represents $\theta_{A_{N-1}}^{2}(\tau)$
in terms of Kapranov's theta functions.
Unfortunately, our attempt failed for some lower $N$'s.
But surprisingly enough, we have found some beautiful formulas that represent 
$\theta_{A_{N-1}}^{2}(\tau)$ in terms of Kapranov's theta functions
with $d=N$:
\begin{conj}
For odd $N\ge 3 $, we can express $\theta_{A_{N-1}}^2(\tau)$ 
as a linear combination of  Kapranov's theta functions
with $d=N$:
\begin{equation}
\theta_{A_{N-1}}^2(\tau)=a_1\theta_{(N)}^{\beta_1}(\tau)+a_2\theta_{(N)}^{\beta_2}(\tau)+\cdots+
a_k\theta_{(N)}^{\beta_k}(\tau),
\end{equation}
where 
\begin{equation}
\theta^{\beta_j}_{(N)}(\tau):=\sum_{m\in {\bf Z}^{N-1}} 
q^{\frac{N}{2}{}^t(m+v(\beta_j))A_{N-1}(m+v(\beta_j))},
\end{equation}
\begin{equation}
v(\beta_j):=\frac{1}{N}A_{N-1}^{-1}
\left(\begin{array}{c}
\beta_j^{(1)}\\
\beta_j^{(2)}\\
\vdots\\
\beta_j^{(N-1)}
\end{array}\right),
\;\;\beta_j^{(n)}\in {\bf Z}.
\end{equation}
Moreover, $a_j $ is a non-negative integer. 
\end{conj}
Using this conjecture, we can describe $ \theta_{A_{N-1}}^2(\tau)$
by Kapranov-type theta function associated with ${\cal O}(-N) $ curve
blow-up formula.
Here we briefly explain the reason why we call $\theta^{\beta_j}_{(N)}(\tau)$
Kapranov-type.
Using the notation of Kapranov,
$\theta^{\beta_j}_{(N)}(\tau)$ can be expressed as, 
\beq
\theta^{\beta_j}_{(N)}(\tau)=\sum_{a\in L}
q^{-\Psi(f,f)/2N+\Psi(a,f)-N\Psi(a,a)/2}\;\;(f=N\cdot v(\beta_{j})),
\enq
and this expression is the same theta function as the one in (\ref{kap})
with $d=N$, except for the factor $q^{-\Psi(f,f)/2N}$ independent of 
$a\in L$.
Notice here that
Kapranov-type theta function is nothing but the 
level $N$ theta function \cite{kac}.

In summary, we have found that the r.h.s. of (\ref{Nth2}) is 
the sum of the ${\cal O}(-N)$ curve
blow-up formulas. Appearance of  ${\cal O}(-N) $ curve blow-up formulas 
instead of ${\cal O}(-2)$ curve blow-up formulas seems to be natural
from the point of view of the results of Nakajima \cite{naka, vafa-witten},
that suggests the appearance of level $N$ representation of affine 
Lie algebra in ${\cal N}=4$ SYM theory with 
gauge group $U(N)$ on ALE spaces.

\subsection{Examples: $N=3,5,7$ Case}
In this subsection, we give some explicit examples of Conjecture 1, 
found by use of Maple V.
\paragraph{$N=3 $ Case}
\begin{equation}
\theta_{A_2}^2(\tau)=\theta_{(3)}^{(1,0)}(\tau)+\theta_{(3)}^{(2,0)}(\tau)+
\theta_{i(3)}^{(4,0)}(\tau),
\end{equation}
where 
\begin{equation}
\theta_{(3)}^{(k,l)}(\tau)=\sum_{m\in{\bf Z}^2}q^{\frac{3}{2}{}^t(m+v(k,l))A_2(m+v(k,l))},
\end{equation}
\begin{equation}
v(k,l)=\frac{1}{3}A_2^{-1}\left(\begin{array}{c}
k\\
l
\end{array}
\right).
\label{3}
\end{equation}
In the following, we give explicit expressions of the theta functions
$\theta_{(3)}^{(1,0)}(\tau),\theta_{(3)}^{(2,0)}(\tau),\theta_{(3)}^{(4,0)}(\tau)$:
\begin{equation}
\theta_{(3)}^{(1,0)}(\tau)=\sum_{(m_1,m_2)\in{\bf Z}^2}
q^{\frac{1}{9}+m_1+3(m_1^2+m_2^2-m_1m_2)}
=: q^{\frac{1}{9}}{\tilde \theta_{(3)}^{0}}(\tau),
\end{equation}
\begin{equation}
\theta_{(3)}^{(2,0)}(\tau)=\sum_{(m_1,m_2)\in{\bf Z}^2}
q^{\frac{4}{9}+2m_1+3(m_1^2+m_2^2-m_1m_2)}
=: q^{\frac{4}{9}}{\tilde\theta_{(3)}^{1}}(\tau),
\end{equation}
\begin{equation}
\theta_{(3)}^{(4,0)}(\tau)=\sum_{(m_1,m_2)\in{\bf Z}^2}
q^{\frac{16}{9}+4m_1+3(m_1^2+m_2^2-m_1m_2)}
=: q^{\frac{7}{9}}{\tilde\theta_{(3)}^{2}}(\tau).
\end{equation}
For later use, we introduced 
${\tilde\theta_{(3)}^{j}}(\tau),(j=0,1,2) $, which
have integral $q$ expansions.
We can easily see that the right hand side of (\ref{3})
consists of 3 types of theta functions, which are classified by
top $q$ powers modulo ${\bf Z}$. 
 We expect that this is the general property 
of Conjecture 1
for odd $N$, which was confirmed by computer check in the $N=5,7$ cases.  

\paragraph{$N=5 $ Case}

\begin{eqnarray}
\theta_{A_4}^2(\tau)
&= &\left.5\theta_{(5)}^{(1,1,1,1)}(\tau)
\right\}=: q{\tilde\theta_{(5)}^{4}}(\tau) 
\nonum
\\
&+ &
\left.\theta_{(5)}^{(1,0,0,1)}(\tau)+2\theta_{(5)}^{(3,1,0,0)}(\tau)+
2\theta_{(5)}^{(1,3,1,0)}(\tau)\right\}=: q^{\frac{1}{5}}
{\tilde\theta_{(5)}^{0}}(\tau)
\nonum
\\
&+ &
\left.\theta_{(5)}^{(2,0,0,2)}(\tau)+2\theta_{(5)}^{(6,2,0,0)}
(\tau)+2\theta_{(5)}^{(2,6,2,0)}(\tau)\right\}=: q^{\frac{4}{5}}
{\tilde\theta_{(5)}^{3}}(\tau)
\nonum
\\
&+ &
\left.\theta_{(5)}^{(0,1,1,0)}(\tau)+2\theta_{(5)}^{(3,0,1,1)}(\tau)+
2\theta_{(5)}^{(1,0,3,0)}(\tau)
\right\}=: q^{\frac{2}{5}}{\tilde\theta_{(5)}^{1}}(\tau)
\nonum
\\
&+& 
\left.
\theta_{(5)}^{(0,2,2,0)}(\tau)+2\theta_{(5)}^{(6,0,2,2)}
(\tau)+2\theta_{(5)}^{(2,0,6,0)}(\tau)\right\}=: 
q^{\frac{3}{5}}{\tilde\theta_{(5)}^{2}}(\tau),
\label{5}
\end{eqnarray}
where 

\begin{equation}
\theta_{(5)}^{(n,k,l,p)}(\tau)=\sum_{m}q^{\frac{5}{2}{}^t(m+v(n,k,l,p))
A_4(m+v(n,k,l,p))},
\end{equation}
\begin{equation}
v(n,k,l,p)=\frac{1}{5}A_4^{-1}
\left(\begin{array}{c}
n\\
k\\
l\\
p
\end{array}
\right).
\end{equation}
As we have already mentioned,
the right hand side of (\ref{5})
has $5$ types of theta functions, which are classified by
top $q$ powers modulo ${\bf Z}$.
In contrast to the $N=3$ case, 
${\tilde \theta_{(5)}}^j(\tau)\;\;(j=0,\cdots,4) $
is expressed by the sum of several theta functions.
Now, we summarize our observations on these common properties in 
the following conjecture:
\begin{conj}
For odd $N\ge 3$, the summands in Conjecture 1 are  
classified by the $q$-powers modulo ${\bf Z}$ as follows:   
\beq
\theta_{A_{N-1}}^2(\tau)=q^{t_N}{\tilde\theta_{(N)}^{0}}(\tau)
+q^{t_N+\frac{1}{N}}{\tilde\theta_{(N)}^{1}}(\tau)+\cdots+
q^{t_N+\frac{N-1}{N}}{\tilde\theta_{(N)}^{N-1}}(\tau),
\enq
where
\beq
t_N=\frac{1}{2N^2}{}^t\rho A_{N-1}\rho=\frac{1}{24}\frac{N^2-1}{N},
\enq
and each ${\tilde\theta_{(N)}^{j}}(\tau)$ has integral $q$ expansions.
Moreover, for $j=0,\cdots,N-1$,
\beq
q^{t_N+\frac{j}{N}}{\tilde\theta_{(N)}^{j}}(\tau)=
a_1^j\theta_{(N)}^{\beta_1^j}(\tau)+
a_2^j\theta_{(N)}^{\beta_2^j}(\tau)+\cdots+
a_{k_{j}}^j\theta_{(N)}^{\beta_{k_{j}}^j}(\tau),
\enq
and $\{ a_p^j \},\;\;(j=0,\cdots,N-1) $ satisfy
\beq
\sum_{p=1}^{k_{0}}a_p^0=
\sum_{p=1}^{k_{1}}a_p^1=\cdots =\sum_{p=1}^{k_{N-1}}a_p^{N-1}=: S_N.
\enq
\end{conj}
The latter half of the above conjecture means that
each ${\tilde\theta_{(N)}^{j}}(\tau)$ consists of sets of 
Kapranov's theta functions
satisfying the property that the sum of coefficients is equal to a fixed 
integer $S_N$.
For example, $S_3=1$ and $S_5=5$.
Even for the $N=7$ case, Conjecture 2 is true and it played the role 
of powerful guide line to find the explicit decomposition.

\paragraph{$N=7 $ Case}

\begin{eqnarray}
\theta_{A_6}^2(\tau)
&=&
%0
\left.\begin{array}{l}
7\theta_{(7)}^{(2,1,1,0,0,0)}(\tau)+7\theta_{(7)}^{(4,2,2,0,0,0)}(\tau)+7\theta_{(7)}^{(6,3,3,0,0,0)}(\tau)
\\
+21\theta_{(7)}^{(1,1,1,1,1,1)}(\tau)+7\theta_{(7)}^{(0,3,1,0,1,0)}(\tau)
\end{array}
\right\}=: q{\tilde \theta_{(7)}^5}(\tau)
%\tag{$q^0$}
\nonum
\\
&& 
%2
\left.
\begin{array}{l}
+\theta_{(7)}^{(0,1,0,0,1,0)}(\tau)+2\theta_{(7)}^{(2,2,0,2,0,0)}(\tau)+2\theta_{(7)}^{(5,0,1,0,0,1)}(\tau)+2\theta_{(7)}^{(0,5,0,1,0,0)}(\tau)
\\
+2\theta_{(7)}^{(1,0,5,0,1,0)}(\tau)+4\theta_{(7)}^{(0,2,2,1,0,0)}(\tau)
+4\theta_{(7)}^{(2,0,2,2,1,0)}(\tau)+4\theta_{(7)}^{(1,2,2,0,2,0)}(\tau)
\\
+6\theta_{(7)}^{(1,1,1,0,3,0)}(\tau)+8\theta_{(7)}^{(0,1,1,1,1,0)}(\tau)+14\theta_{(7)}^{(1,1,1,1,0,3)}(\tau)
\end{array}
\right\}
=: q^{\frac{2}{7}}{\tilde \theta_{(7)}^0}(\tau)
%\tag{$q^{\frac{2}{7}}$}
\nonum
\\
&&
%1
\left.
\begin{array}{l}
+\theta_{(7)}^{(0,2,0,0,2,0)}(\tau)+2\theta_{(7)}^{(4,4,0,4,0,0)}(\tau)+2\theta_{(7)}^{(10,0,2,0,0,2)}(\tau)+2\theta_{(7)}^{(0,10,0,2,0,0)}(\tau)
\\
+2\theta_{(7)}^{(2,0,10,0,2,0)}(\tau)+4\theta_{(7)}^{(0,4,4,2,0,0)}(\tau)
+4\theta_{(7)}^{(4,0,4,4,2,0)}(\tau)+4\theta_{(7)}^{(2,4,4,0,4,0)}(\tau)
\\
+6\theta_{(7)}^{(2,2,2,0,6,0)}(\tau)+8\theta_{(7)}^{(0,2,2,2,2,0)}(\tau)+14\theta_{(7)}^{(2,2,2,2,0,6)}(\tau)
\end{array}
\right\}
=: q^{\frac{8}{7}}{\tilde \theta_{(7)}^6}(\tau)
%\tag{$q^{\frac{1}{7}}$}
\nonum
\\
&&
%4
\left.
\begin{array}{l}
+\theta_{(7)}^{(0,3,0,0,3,0)}(\tau)+2\theta_{(7)}^{(6,6,0,6,0,0)}(\tau)+2\theta_{(7)}^{(15,0,3,0,0,3)}(\tau)+2\theta_{(7)}^{(0,15,0,3,0,0)}(\tau)
\\
+2\theta_{(7)}^{(3,0,15,0,3,0)}(\tau)+4\theta_{(7)}^{(0,6,6,3,0,0)}(\tau)
+4\theta_{(7)}^{(6,0,6,6,3,0)}(\tau)+4\theta_{(7)}^{(3,6,6,0,6,0)}(\tau)
\\
+6\theta_{(7)}^{(3,3,3,0,9,0)}(\tau)+8\theta_{(7)}^{(0,3,3,3,3,0)}(\tau)+14\theta_{(7)}^{(3,3,3,3,0,9)}(\tau)
\end{array}
\right\}
=: q^{\frac{4}{7}}{\tilde \theta_{(7)}^2}(\tau)
%\tag{$q^{\frac{4}{7}}$}
\nonum
\\
&&
%3
\left.
\begin{array}{l}
+\theta_{(7)}^{(0,0,1,1,0,0)}(\tau)+2\theta_{(7)}^{(4,0,1,0,0,0)}(\tau)+2\theta_{(7)}^{(2,4,0,1,0,0)}(\tau)+2\theta_{(7)}^{(1,0,0,5,0,0)}(\tau)
\\
+3\theta_{(7)}^{(2,1,0,0,1,2)}(\tau)+4\theta_{(7)}^{(2,0,1,1,1,0)}(\tau)
+4\theta_{(7)}^{(2,1,2,1,0,0)}(\tau)+4\theta_{(7)}^{(1,2,1,2,1,0)}(\tau)
\\
+5\theta_{(7)}^{(1,0,2,2,0,1)}(\tau)+6\theta_{(7)}^{(1,2,1,0,0,1)}(\tau)+8\theta_{(7)}^{(1,1,0,2,2,0)}(\tau)+8\theta_{(7)}^{(2,2,0,1,1,1)}(\tau)
\end{array}
\right\}
=: q^{\frac{3}{7}}{\tilde \theta_{(7)}^1}(\tau)
%\tag{$q^{\frac{3}{7}}$}
\nonum
\\
&&
%5
\left.
\begin{array}{l}
+\theta_{(7)}^{(0,0,2,2,0,0)}(\tau)+2\theta_{(7)}^{(8,0,2,0,0,0)}(\tau)+2\theta_{(7)}^{(4,8,0,2,0,0)}(\tau)+2\theta_{(7)}^{(2,0,0,10,0,0)}(\tau)
\\
+3\theta_{(7)}^{(4,2,0,0,2,4)}(\tau)+4\theta_{(7)}^{(4,0,2,2,2,0)}(\tau)
+4\theta_{(7)}^{(4,2,4,2,0,0)}(\tau)+4\theta_{(7)}^{(2,4,2,4,2,0)}(\tau)
\\
+5\theta_{(7)}^{(2,0,4,4,0,2)}(\tau)+6\theta_{(7)}^{(2,4,2,0,0,2)}(\tau)+8\theta_{(7)}^{(2,2,0,4,4,0)}(\tau)+8\theta_{(7)}^{(4,4,0,2,2,2)}(\tau)
\end{array}
\right\}
=: q^{\frac{5}{7}}{\tilde \theta_{(7)}^3}(\tau)
%\tag{$q^{\frac{5}{7}}$}
\nonum
\\
&&
%6
\left.
\begin{array}{l}
+\theta_{(7)}^{(0,0,3,3,0,0)}(\tau)+2\theta_{(7)}^{(12,0,3,0,0,0)}(\tau)+2\theta_{(7)}^{(6,12,0,3,0,0)}(\tau)+2\theta_{(7)}^{(3,0,0,15,0,0)}(\tau)
\\
+3\theta_{(7)}^{(6,3,0,0,3,6)}(\tau)+4\theta_{(7)}^{(6,0,3,3,3,0)}(\tau)
+4\theta_{(7)}^{(6,3,6,3,0,0)}(\tau)+4\theta_{(7)}^{(3,6,3,6,3,0)}(\tau)
\\
+5\theta_{(7)}^{(3,0,6,6,0,3)}(\tau)+6\theta_{(7)}^{(3,6,3,0,0,3)}(\tau)+8\theta_{(7)}^{(3,3,0,6,6,0)}(\tau)+8\theta_{(7)}^{(6,6,0,3,3,3)}(\tau)
\end{array}
\right\}
=: q^{\frac{6}{7}}{\tilde \theta_{(7)}^4}(\tau),
%\tag{$q^{\frac{6}{7}}$}
\nonum
\\
\end{eqnarray}
where 

\begin{equation}
\theta_{(7)}^{(n,k,l,p,r,s)}(\tau)=\sum_{m}q^{\frac{7}{2}{}^t(m+v(n,k,l,p,r,s))A_6(m+v(n,k,l,p,r,s))},
\end{equation}
\begin{equation}
v(n,k,l,p,r,s)=\frac{1}{7}A_6^{-1}\left(\begin{array}{c}
n\\
k\\
l\\
p\\
r\\
s
\end{array}
\right).
\end{equation}
This computation  was very hard because $N=7$ is critical bound for the 
power of our computer. However, Conjecture 2 is also true
with  $S_7=49$. 
Here, we have to remark on a subtle point.
In this case, decomposition of ${\tilde \theta_{(7)}^j}(\tau)$ 
into $\theta_{(7)}^{\beta_j}(\tau)$'s is not unique 
because there are some  non-trivial identities 
between $\theta_{(7)}^{\beta_j}(\tau)$'s.
Up to now, we have confirmed our conjecture up to $N\leq 7$ 
case, but we believe that Conjecture 2 is also true for odd $N> 7$ even 
if $N$ is not prime. 

\section{Final Formulas}
\label{sec:5}
\setcounter{equation}{0}
Finally, we generalize our previous work on $SU(2)/{\bf Z}_2$
partition function on orbifold-$T^4/{\bf Z}_2$ \cite{jin}
to $SU(N)/{\bf Z}_N$ case,
using the formulas prepared in Sec. 3,4.
We choose the following processes.
First we introduce $K3$ partition function $Z^{K3}$, 
and then separate $Z^{K3}$ into $Z^{S_0}$ (the contribution from $S_0$)
and $Z^{B}$ (the blow-up formula).
As a concrete example,
we choose the partition function of $SU(N)/{\bf Z}_N$ theory on $K3$ surface 
with {}'t Hooft flux $v^{2}=0 \;\;(mod ~N)$. 
Generalization to other cases is obviously straightforward.
Using the same method as the one in Sec. 2, the partition function  
is given by, 
\begin{eqnarray}
Z^{K3}_{0}(\tau)
&=&
\frac{1}{N}\left(
\frac{1}{\eta^{24}(\frac{\tau}{N})}
+\frac{1}{\eta^{24}(\frac{\tau+1}{N})}
+\cdots
+\frac{1}{\eta^{24}(\frac{\tau+N-1}{N})}
\right)
\nonum
\\
&= &
\frac{1}{N}\left(
\frac{1}{\eta^{8}(\frac{\tau}{N})}\frac{1}{\eta^{16}(\frac{\tau}{N})}
+\frac{1}{\eta^{8}(\frac{\tau+1}{N})}\frac{1}{\eta^{16}(\frac{\tau+1}{N})}
+\cdots
+\frac{1}{\eta^{8}(\frac{\tau+N-1}{N})}\frac{1}{\eta^{16}(\frac{\tau+N-1}{N})}
\right)
\nonum
\\
&=&
Z^{S_0}_0(\tau)Z^{B}_1(\tau)
+Z^{S_0}_1(\tau)Z^{B}_0(\tau)
+\cdots
+Z^{S_0}_{N-1}(\tau)Z^{B}_2(\tau).
\label{f1}
\end{eqnarray}
Here $Z^{S_0}_j(\tau)$ is already computed in Sec.2 as follows:
\begin{eqnarray}
Z^{S_0}_j(\tau)
&=&
\frac{1}{N}\left(
\frac{1}{\eta^8(\frac{\tau}{N})}
+\zeta^{\frac{1}{3}}_N\zeta^{-j}_N\frac{1}{\eta^8(\frac{\tau+1}{N})}
+\cdots
+\zeta^{\frac{N-1}{3}}_N\zeta^{-j(N-1)}_N\frac{1}{\eta^8(\frac{\tau+N-1}{N})}
\right)
\nonum
\\
&=&
q^{-\frac{1}{3N}}(q^{\frac{j}{N}}+\cdots).
\label{f2}
\end{eqnarray}
From (\ref{f1}) and (\ref{f2}), we can pick up $Z^B_j(\tau)$, the 
contribution from blowing up the 16 singularities:
\begin{eqnarray}
Z^{B}_j(\tau)
&=&
\frac{1}{N}\left(
\frac{1}{\eta^{16}(\frac{\tau}{N})}
+\zeta^{\frac{2}{3}}_N\zeta^{-j}_N\frac{1}{\eta^{16}(\frac{\tau+1}{N})}
+\cdots
+\zeta^{\frac{2(N-1)}{3}}_N\zeta^{-j(N-1)}_N\frac{1}{\eta^{16}(\frac{\tau+N-1}{N})}
\right)
\nonum
\\
&=&
q^{-\frac{2}{3N}}(q^{\frac{j}{N}}+\cdots).
\label{f3}
\end{eqnarray}
Using the key identities and conjecture 2, 
we can rewrite (\ref{f3}) in terms of blow-up formulas as follows:  
\begin{eqnarray}
Z^{B}_j(\tau)
&=&
\frac{1}{N}
\frac{1}{\eta^{16N}(\tau)}
\left(
(\theta^{2}_{A_{N-1}}(\tau))^{16}
+\zeta_N^{-j}(\theta^{2}_{A_{N-1}}(\tau+1))^{16}
+\cdots
+\zeta^{-j(N-1)}(\theta^{2}_{A_{N-1}}(\tau+N-1))^{16}
\right)
\nonum
\\
&=&
\frac{1}{N}
\frac{1}{\eta^{16N}(\tau)}q^{\frac{2}{3}\frac{N^2-1}{N}}\left[\left(
\sum_{j=0}^{N-1}q^{\frac{j}{N}}{\tilde \theta_{(N)}^{j}(\tau)}
\right)^{16}\right]_{q^{\frac{j}{N}}},
\label{f4}
\end{eqnarray}
where $[\cdots] _{q^{\frac{j}{N}}}$ stands 
for the operation of 
projecting out $q^{\frac{j}{N}+n}\;\;(n\in {\bf Z}_{\geq 0})$ powers.
Here we will give explicit expression of (\ref{f4}) in the $N=3$ case:

\begin{eqnarray}
Z^{B}_0(\tau)
&=&
\frac{1}{3}
\frac{1}{\eta^{48}(\tau)}
\left(
(\theta^{2}_{A_{2}}(\tau))^{16}
+(\theta^{2}_{A_{2}}(\tau+1))^{16}
+(\theta^{2}_{A_{2}}(\tau+2))^{16}
\right)
\nonum
\\
&=&
\frac{1}{3}
\frac{1}{\eta^{48}(\tau)}q^{\frac{16}{9}}\left[\left(
{\tilde \theta_{(3)}^{0}(\tau)}
+q^{\frac{1}{3}}{\tilde \theta_{(3)}^{1}(\tau)}
+q^{\frac{2}{3}}{\tilde \theta_{(3)}^{2}(\tau)}
\right)^{16}\right]_{q^{\frac{0}{3}}}
\nonum
\\
&=&
\frac{1}{3}
\frac{1}{\eta^{48}(\tau)}q^{\frac{16}{9}}\left(
({\tilde \theta_{(3)}^{0}(\tau)})^{16}
+560q({\tilde \theta_{(3)}^{0}(\tau)})^{13}({\tilde \theta_{(3)}^{1}(\tau)})^3
+\cdots+16q^{10}({\tilde \theta_{(3)}^{0}(\tau)})({\tilde \theta_{(3)}^{2}(\tau)})^{15}
\right).\nonum
\\
&&
\end{eqnarray}
In this way, we have reconstructed $K3$ partition function which has
the factor of the partition function on $S_0$ and that
of the blow-up formulas.
Note that our processes in this section are opposite direction 
to our previous work \cite{jin}.
In the previous work on $SU(2)$ gauge group,
we first prepared the partition function on $S_0$
and ${\cal O}(-2)$ curve blow-up formulas.
Next,  we multiplied and summed up these factors 
so that the total partition function satisfy the $S$-duality conjecture.
The final result was surely the $K3$ partition function 
derived  by Vafa and Witten.
In both cases, the expression of the partition function we have obtained 
has the same structure, i.e.,
it consists of 
the factor of the partition function on $S_0$ and that
of the blow-up formulas.
But we have to notice that
the contribution from the blow-up formulas can be expressed
by ${\cal O }(-N)$ blow-up formulas with $\frac{j}{N}+{\bf Z}_{\geq 0}$ 
powers. Of course, this result 
still needs appropriate geometrical interpretation.
\section{Conclusion}
\label{sec:6}
In this paper, we reconstructed the $K3$ partition function 
of ${\cal N}=4$ $SU(N)/{\bf Z}_N$ gauge theory
using orbifold construction $T^4/{\bf Z}_2$.
The key point lies on the fact that 
we can rewrite the function $1/\eta(\frac{\tau}{N})$ in terms of 
${\cal O}(-N)$ curve blow-up formulas.

The remaining most serious problem is
geometrical interpretation of ${\cal O}(-N)$ curve blow-up formulas 
in our theory.
Since coefficients of ${\cal O}(-N)$ curve blow-up formulas
are all integers,
we hope that there may be some nice geometrical interpretation
similar to  the one in the $SU(2)$ case. 

 From the point of view of the work of Nakajima \cite{naka}, 
we have to say that, for the $SU(N)$ case, 
the contribution from blow-up comes from
$U(N)$ gauge theory on ALE $A_{N-1}$ space.
Then appearance of level $N$ $SU(N)$ theta functions 
seems to be natural.
If we believe these scenario, we can expect 
the following interesting applications of our result.
According to \cite{naka}, there are already $ADE$ gauge theory on
ALE space associated with the same $ADE$ Lie algebra.
Then, we can speculate the form of  "$ADE$ blow-up" formula 
using the denominator identity of $ADE$ affine Lie algebra 
as we did in Sec. 3 of this paper.
Moreover, we may be able 
to determine the form of the partition function of ${\cal N}=4$
$ADE$ gauge theory on $K3$ surfaces, by tracing back the process of our 
computation in this paper.

The most crucial point in this paper is the  
process of rewriting the function $1/\eta(\frac{\tau}{N})$
in terms of ${\cal O}(-N)$ curve blow-up formula.
This process seems to suggest existence of a kind of 
duality between the bulk contribution and the blow-up contribution
in the sense of Vafa and Witten \cite{vafa-witten}.
Physical meaning of this duality remains a problem to be cleared up.
We expect that we may estimate the form of the partition function of 
$ADE$ gauge theory on $K3$,
if we apply the formula to the "$ADE$ blow-up formula". 

Finally, we point out level-rank duality in affine Lie algebra, that 
states duality between $\hat{sl}(l)_{r}$ and $\hat{sl}(r)_{l}$ \cite{NT}. 
In our case, we observed appearance of $\hat{sl}(N)_{N}$, 
which is self-dual in the context of level-rank duality. We expect that
this special symmetry may have some physical meaning. 

{\bf Acknowledgment}\\
We would like to thank Prof. K.Yoshioka for useful discussion
at the early stage of this work. We also thank Prof. N.~Kawamoto for 
carefully reading our manuscript. 
T.S. would like to thank the members at the Elementary Particle
Group of Dept. of Phys. of Univ. of Hokkaido 
for consulting on the computers. M.J. would like to thank Dr. M. Naka 
and Prof. A. Nakayashiki for discussions. Research of M.J. is partially 
supported by a grant of Japan Society for Promotion of Science.


\begin{thebibliography}{99}


\bibitem{Asp} P.S.Aspinwall. {\it K3 Surfaces and String Duality},
hep-th/9611137.
\bibitem{bonelli} G.Bonelli. {\it The geometry of M5-branes and TQFTs},
hep-th/0012075.
\bibitem{dijkgraaf} R. Dijkgraaf, J.-S. Park, B. Schroers.
{\it N=4 Supersymmetric Yang-Mills Theory on a K\"ahler Surface},
 hep-th/9801066 .
\bibitem{Fukaya}K.Fukaya.  {\it Topology, geometry and field theory}
, World Scientific 1994.
\bibitem{jin} M.Jinzenji and T.Sasaki.
 {\it $N=4$ Supersymmetric Yang-Mills Theory on Orbifold-$T^4/{\bf Z_2}$},
Mod.Phys.Lett. A16 (2001) 411-428.
\bibitem{kac} V. G. Kac.
 {\it Infinite dimensional Lie algebras},
 Cambridge 1990.
\bibitem{kap} M.Kapranov.
 {\it The Elliptic Curve in the $S$-Duality Theory and Eisenstein Series for
Kac-Moody Groups},
 math.AG/0001005.
\bibitem{laba}
J. M. F. Labastida, Carlos Lozano. 
 {\it  Mathai-Quillen Formulation of Twisted N=4 Supersymmetric 
Gauge Theories in Four Dimensions},
 Nucl.Phys. B 502 (1997) 741. 
\bibitem{lozano}
J. M. F. Labastida, Carlos Lozano.
 {\it  The Vafa-Witten Theory for Gauge Group SU(N)},
 hep-th/9903172. 
\bibitem{qin}W-P.Li and  Z.Qin. {\it On blow-up formulae for the S-duality 
conjectures of Vafa and Witten},
 Invent. Math. 136 (1999), no. 2, 451-482.
\bibitem{mac}I.G.Macdonald.
 {\it Affine root systems and Dedekind's $\eta$-functions},
 Invent. math. {\bf 15} (1972), 91-143.
\bibitem{m-v}J. A. Minahan, D. Nemeschansky, C. Vafa, N. P. Warner.
 {\it E-Strings and N=4 Topological Yang-Mills Theories},
 Nucl.Phys. B 527 (1998) 581-623.
\bibitem{NT} T. Nakanishi, A. Tsuchiya.
{\it Level-Rank Duality of WZW Models in Conformal Field Theory},
Commun. Math. Phys. 144, 351-372 (1992). 
\bibitem{M-O}
C.Montonen and D.Olive.
{\it Magnetic monopoles as gauge particles ?}
  Phys.Lett.B 72 (1977) 117;\\
P.Goddard, J.Nyuts and D.Olive.
{\it Gauge theories and magnetic charge},
 Nucl. Phys. B125 (1977) 1.
\bibitem{mukai}
S.Mukai. 
{\it Symplectic structure of the moduli space of sheves on an abelian or K3
surface}, Inv.Math. 77 (1984) 101;\\
L.G\"ottsche.
{\it The Betti numbers of the Hilbert scheme of points on a smooth 
projective surface},
 Math.Ann. 286 (1990) 193.
\bibitem{nak}
H.Nakajima.
{\it Lectures on Hilbert schemes of points on surfaces},
to appear.
\bibitem{naka}
H.Nakajima.
{\it Instantons on ALE spaces, quiver varieties, and Kac-Moody algebras},
Duke Math. J. 76 (1994);\\
{\it Gauge theory on resolutions of simple singularities and simple Lie algebras.}, Internat. Math. Res. Notices 1994.
\bibitem{sako} A. Sako, T. Sasaki.
{\it Euler number of Instanton Moduli space and Seiberg-Witten invariants},
J.Math.Phys. 42 (2001) 130-157.
\bibitem{vafa-witten} C.Vafa and E.Witten.
{\it A strong coupling test of S-duality}, 
Nucl. Phys. B431 (1994) 3.
\bibitem{witten}E.Witten. 
{\it Supersymmetric Yang-Mills theory on a four manifold},
J. Math. Phys. 35 (1994) 5101-5135.
\bibitem{yoshioka}K.Yoshioka. {\it The Betti numbers of the moduli space of 
stable sheaves of rank $2$ on ${\bf P}^2$}, 
J. reine angew. Math. {\bf 453} (1994), 193-220.
\bibitem{yoshi}
K.Yoshioka.
{\it Euler characteristics of SU(2) instanton moduli spaces 
on rational elliptic surfaces},
Commun.Math.Phys. 205 (1999) 501-517.
\bibitem{yoshiell} K.Yoshioka. {\it Numbers of ${\bf F}_{q}$-rational 
points of moduli of stable sheaves on elliptic surfaces},
moduli of vector bundles, Lect. Notes in Pure and Applied Math. 179, 297-305, 
Marcel Dekker.
\bibitem{yoshi2}K.Yoshioka. {\it Moduli spaces of stable sheaves on 
abelian surfaces}, \\
math-AG/0009001.
\bibitem{yoshihecke}K.Yoshioka. {\it Irreducibility of moduli spaces of vector bundles on K3 surfaces},\\
math.AG/9907001. 
\bibitem{yoshisun}K.Yoshioka.
{\it Betti numbers of the moduli space of 
stable sheaves on some surfaces},
Nucl.Phys.B (Proc.Suppl.) 46 (1996) 263-268.
\bibitem{prime}K.Yoshioka.
{\it Some notes on the moduli of stable sheaves on elliptic surfaces},
Nagoya Math.J.154 (1999), 73-102.
\bibitem{priv}K.Yoshioka. {\it private communication.}
\end{thebibliography}
\end{document}